\documentclass[aps,prl,nofootinbib,preprintnumbers,twocolumn,superscriptaddress,floatfix]{revtex4}
\usepackage{amsmath,amssymb}
\usepackage[dvips, pdftex]{graphicx}
\usepackage[tight,footnotesize]{subfigure}
\usepackage[usenames]{color}
\usepackage[dvipsnames]{xcolor}
\usepackage{float}
\usepackage{wrapfig}
\usepackage{hyperref}
\usepackage{mathtools}
\usepackage{physics}
\usepackage[utf8]{inputenc}
\usepackage{graphicx}
\graphicspath{ {./images/} }

\begin{document}

\def\ga{\mathrel{\raise.3ex\hbox{$>$\kern-.75em\lower1ex\hbox{$\sim$}}}}
\def\la{\mathrel{\raise.3ex\hbox{$<$\kern-.75em\lower1ex\hbox{$\sim$}}}}

\def\be{\begin{equation}}
\def\ee{\end{equation}}
\def\bea{\begin{eqnarray}}
\def\eea{\end{eqnarray}}

\def\betap{\tilde\beta}
\def\del{\delta_{\rm PBH}^{\rm local}}
\def\Msun{M_\odot}
\def\Rcl{R_{\rm clust}}
\def\fPBH{f_{\rm PBH}}

\preprint{KCL-PH-TH-2025-24}

\title{Popcorn in the sky:  \\
Identifying primordial black holes in  the gravitational-wave background}

\author{Eleni Bagui}
\affiliation{Service de Physique Th\'eorique, Universit\'e Libre de Bruxelles (ULB), Boulevard du Triomphe, CP225, 1050 Brussels, Belgium.}

\author{Sebastien Clesse}
\affiliation{Service de Physique Th\'eorique, Universit\'e Libre de Bruxelles (ULB), Boulevard du Triomphe, CP225, 1050 Brussels, Belgium.}

\author{Federico De Lillo}
\affiliation{Universiteit Antwerpen, Prinsstraat 13, 2000 Antwerpen, Belgium}

\author{Alexander C. Jenkins}
\affiliation{Kavli Institute for Cosmology, University of Cambridge, Madingley Road, Cambridge CB3 0HA, UK}
\affiliation{DAMTP, University of Cambridge, Wilberforce Road, Cambridge CB3 0WA, UK}

\author{Mairi Sakellariadou}
\affiliation{Theoretical Particle Physics and Cosmology Group,  Physics Department, \\ King's College London, University of London, Strand, London WC2R 2LS, United Kingdom}

\date{\today}

\begin{abstract}
Primordial black holes (PBHs) are possible sources of a gravitational-wave background (GWB), detectable with the next observing runs of LIGO--Virgo--KAGRA. In case of a detection, it will be crucial to distinguish the possible sources of this GWB. One under-explored possibility is to exploit the duty cycle that quantifies the number of sources present in the time domain signal, which can be very different depending on the nature and population of the sources.
We compute the duty cycle for a realistic population of PBH binaries, isolating the shot-noise, popcorn and continuous contributions to the GWB. We identify the dependence of the duty cycle on the signal frequency, duration and amplitude as a crucial metric for distinguishing PBHs from other sources in the GWB and constraining PBH models.
Our work motivates the development of specific analysis tools to extract these observables, in order to unlock new cosmological insights with upcoming GW data.
\end{abstract}

\maketitle

\textbf{Introduction---} Since the first detection by Advanced LIGO of a black hole (BH) binary merger~\cite{GW150914_LIGOScientific:2016aoc}, more than 200 compact object coalescences have been observed by the LIGO--Virgo--KAGRA (LVK) collaboration~\cite{LIGOScientific:2025slb,LIGOScientific:2025pvj}. These detections have rekindled interest in primordial black holes (PBHs, see~\cite{Carr:2023tpt,LISACosmologyWorkingGroup:2023njw} for recent reviews).  PBHs may have formed in the early Universe~\cite{Zeldovich:1967lct,Hawking:1971ei,Carr:1974nx,Chapline:1975ojl} from the gravitational collapse of inflationary inhomogeneities (see also~\cite{LISACosmologyWorkingGroup:2023njw} for alternative mechanisms), and may constitute a significant fraction of the dark matter.   They can form binaries immediately after formation~\cite{Nakamura:1997sm,Sasaki:2016jop} when two PBHs are initially very close to each other (``early PBH binaries''), or dynamically {at later times} inside dense clusters~\cite{Clesse:2016vqa,Bird:2016dcv} (``late PBH binaries''), with the former believed to dominate in terms of GW emission. Recent analyses seem to favor a combination of PBHs and stellar black holes to explain GW observations~\cite{Escriva:2022bwe,Franciolini:2022tfm}. The detection of subsolar-mass candidates~\cite{Morras:2023jvb,Prunier:2023cyv,Phukon:2021cus,LIGOScientific:2023lpe} such as SSM200308~\cite{Prunier:2023cyv} offer another promising means of testing the PBH hypothesis, but their statistical significance still remains relatively low.

A major milestone for future LVK observations will be the detection of the gravitational-wave background (GWB): a persistent signal composed of many GW sources throughout cosmic time, including (potentially) a variety of early-Universe processes, as well as a large number of unresolved compact binaries, such as stellar black hole and neutron star mergers, and possibly PBH mergers.
The GWB from early PBH binaries was calculated in~\cite{Bagui:2021dqi} for broad PBH mass distributions that include the effects of the QCD phase transition on PBH formation~\cite{Byrnes:2018clq,QCD_mass_distribution_Carr:2019kxo,Escriva:2022bwe,Musco:2023dak}.
This typically boosts the PBH abundance in the solar-mass range, generating a GWB that will be probed by upcoming LVK observing runs.  For this reason, PBHs will be a well-motivated candidate in the case of a GWB detection.
However, the shape of the associated GWB spectrum is highly degenerate with that expected from astrophysical compact binaries, and it is therefore unclear how to disentangle these two possible GWB sources.

In this \textit{Letter} we show that a key tool for tackling this challenge will be the reconstruction of the time-domain \textit{duty cycle}---i.e., the mean number of sources present in the GWB signal at any given time.
This encodes the source population, and, as we show below, allows one to distinguish between different GWB signals with similar frequency spectra. We compute for the first time the duty cycle associated with the dominant GWB from early PBHs binaries, and compare this to expectations for stellar black-hole and neutron-star binaries.
More specifically, we highlight that the so-called \textit{popcorn} regime, where the duty cycle is slightly below unity, is a good discriminator between PBHs and stellar BHs, especially when assessed as a function of the signal duration in each frequency bin and of the strain amplitude.  This combination of information in principle allows one to reconstruct any merger rate distribution as a function of component masses and redshift, enabling powerful tests of PBH models with LVK data.

\textbf{Merger rate of early PBH binaries---}
The GWB from PBHs in the LVK frequency band is dominated by ``early'' binares formed before matter-radiation equality.   The associated merger rate distribution (for ordered masses $m_1 \ge m_2$) can be modelled as~\cite{PBH_rate_Hutsi:2020sol, PBH_rate_Raidal:2018bbj}
\bea
        &\mathcal 
        R (m_1,m_2,z) =
        \frac{1.6 \times 10^6}{\rm Gpc^3 yr} \times f_{\rm sup}(m_1,m_2,z) f_{\rm PBH}^{53/37} \nonumber \\ 
        & \times \phi(m_1)  \phi(m_2)
        \left(\frac{m_1 + m_2}{M_\odot}\right)^{-\frac{32}{37}} \left[\frac{t(z)}{t_0} \frac{m_1 m_2}{(m_1+m_2)^2}\right]^{-\frac{34}{37}} ~,  \label{eq:cosmomerg}
\eea
where $f_{\rm PBH}$ is the fraction of dark matter composed of PBHs, $t_0$ is the age of the Universe, $\phi (m)$ is the normalized PBH mass distribution, and $f_{\rm sup}$~\cite{PBH_rate_Hutsi:2020sol} is a rate suppression factor that takes into account binary perturbations~\cite{PBH_rate_Raidal:2018bbj} by other PBHs, by clusters of PBHs and by matter inhomogeneities.  
This factor can be estimated using N-body simulations of PBH binaries and, when $z \leq 100$ and $f_{\rm PBH} \geq 0.0035$, can be approximated as $f_{\rm sup} \approx 2.3 \times 10^{-3} \times f_{\rm PBH}^{0.65} \times [t(z)/t_0]^{-0.29} $~\cite{Escriva:2022bwe} for a peaked mass distribution. In this paper, we consider the extended mass distributions of~\cite{Byrnes:2018clq,Escriva:2022bwe, QCD_mass_distribution_Musco:2012au} that include the imprints of the QCD epoch when the PBH formation is boosted, leading to a peak at the solar-mass scale, thereby justifying this assumption.  Note that we neglect the complex effects of light perturbers on PBH binaries, which is still an open problem~\cite{Escriva:2022bwe}.

\begin{figure}
    \centering
    \includegraphics[width = \columnwidth]{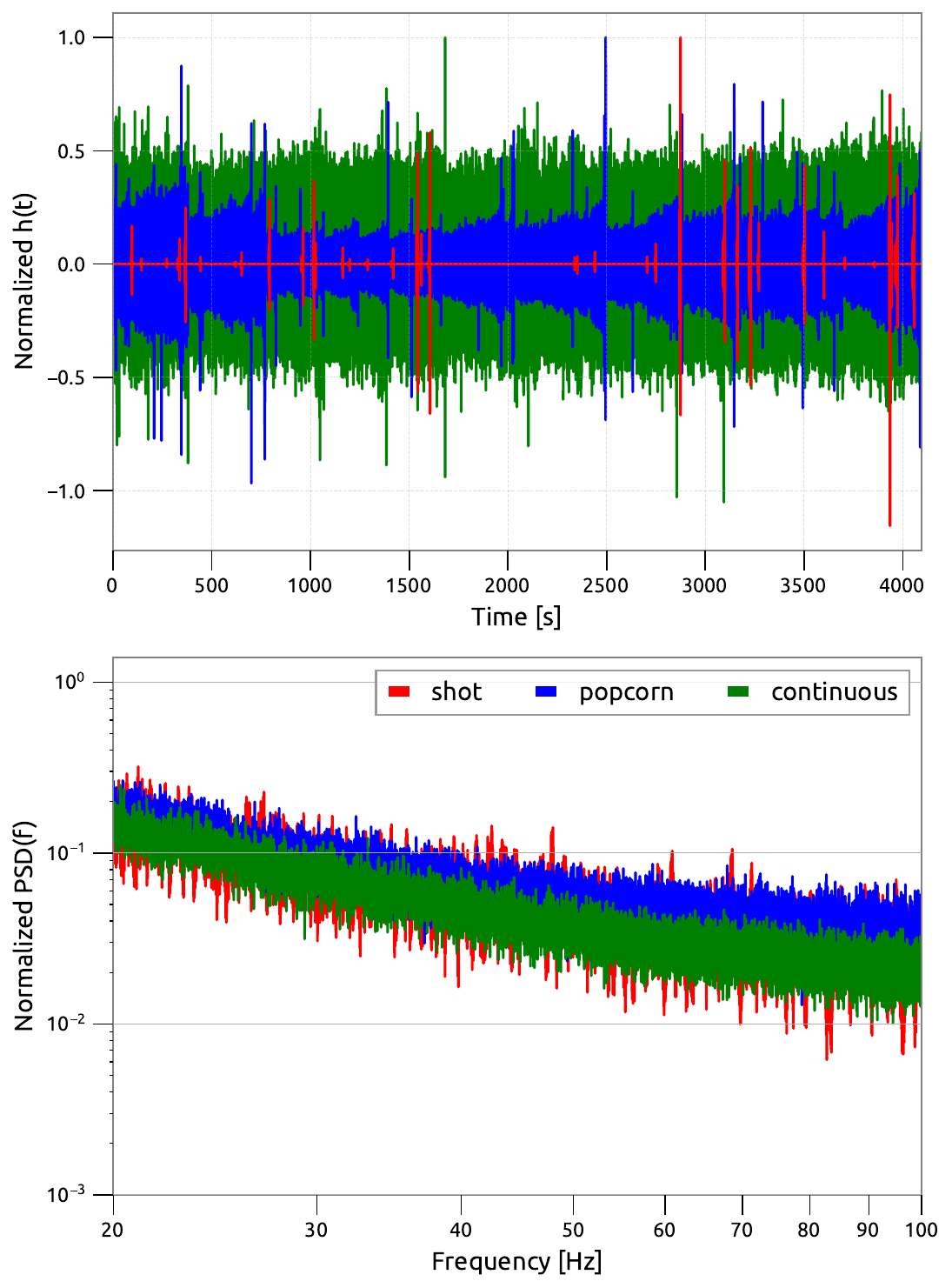}
    \caption{%
    Normalized strain time series and power spectral densities in the shot-noise (red), popcorn (blue), and continuous (green) regimes for three toy models of compact-binary populations.
    Despite having near-identical frequency spectra, the three models are drastically different in the time domain, illustrating the additional information accessible via statistics such as the duty cycle.}
    \label{fig:Popcorn_plot}
\end{figure}

\textbf{Duty cycle---}
GWB signals are typically characterised by their frequency spectra.  However, even for signals with similar spectral properties, their time-domain behaviour may be quite different (cf. Fig.~\ref{fig:Popcorn_plot}), and this behaviour can therefore be leveraged to break model degeneracies.
A key quantity for this purpose is  the \textit{duty cycle} $\Delta$ \cite{Coward:2006df, Regimbau:2009rk, Regimbau:2011rp, Regimbau:2022mdu}, which is the average number of overlapping signals in a given frequency band, at a given time. This quantity, in the observer's frame, is defined as the ratio of the typical duration of a merger signal $\Delta T(f)$ (here defined per unit of logarithmic frequency) to the average time interval between two successive events,
\begin{equation}
    \label{eq:duty_cycle_definition}
    \Delta(z,f) = \int_{\mathbf{\theta}} \int_{0}^{z} \Delta T (z',f,{\mathbf \theta}) \frac{{\rm d}R^{\rm o}(z',{\mathbf \theta})}{{\rm d}z'}{\rm d}z' {\rm d {\mathbf \theta}},
\end{equation}
where $\mathbf{\theta}$ denotes the population parameters, such as the binary component masses $m_1$ and $m_2$. Here, ${\rm d}R^{\rm o}(z') / {\rm d}z' = \mathcal R(z') {\rm d} V(z')/{\rm d} z'$ is the event rate per unit of redshift in the observer's frame and $\mathcal R(z')$ is the GW event rate per comoving volume $V(z)$.

The value of the duty cycle can vary greatly depending on the value of $\Delta T (z')$, given by \cite{Regimbau:2022mdu}
\begin{align}
    \label{eq:duration}
    \Delta T (z',f, \mathcal{M}_{c}) = &\frac{5 c^{5}}{256 \pi^{8/3}G^{5/3}} \, \left[(1+z')\mathcal{M}_c\right]^{-5/3} \nonumber
    \\ &\times \left(f_{\mathrm{L}}^{-8/3} - f_{\mathrm{H}}^{-8/3}\right),
\end{align}
where $\mathcal{M}_{\rm c}$ is the chirp mass of the binary, and $f_{\rm{L}}$, $f_{\rm{H}}$ are the lower and upper limits of the frequency band of interest, respectively. Given a certain frequency $f$, we define the band with a unit logarithmic width, such that $f_{\rm{L}} \equiv f$ and $f_{\rm{H}} \equiv \max\left\{ f_{\rm{L} } , \min \left\{ {\rm e}\,f_{\rm L}, f_{\rm{ISCO}}/(1+z)\right\}\right\}$.  This way, we obtain the signal duration per $e$-fold in frequency as long as $f$ is smaller than the frequency of the innermost stable circular orbit $f_{\rm ISCO} = 4400 \, {\rm Hz} \, M_\odot / (m_1+m_2) $, beyond which the binary leaves the inspiral regime.  We note that the redshift dependence in this ISCO frequency cutoff was overlooked or neglected in previous GWB analyses~\cite{Bagui:2021dqi}. Its impact is significant for this work given the large number of sources at high redshift.

\begin{figure}
\begin{minipage}[t]{0.99\columnwidth}
  \includegraphics[width=1\linewidth]{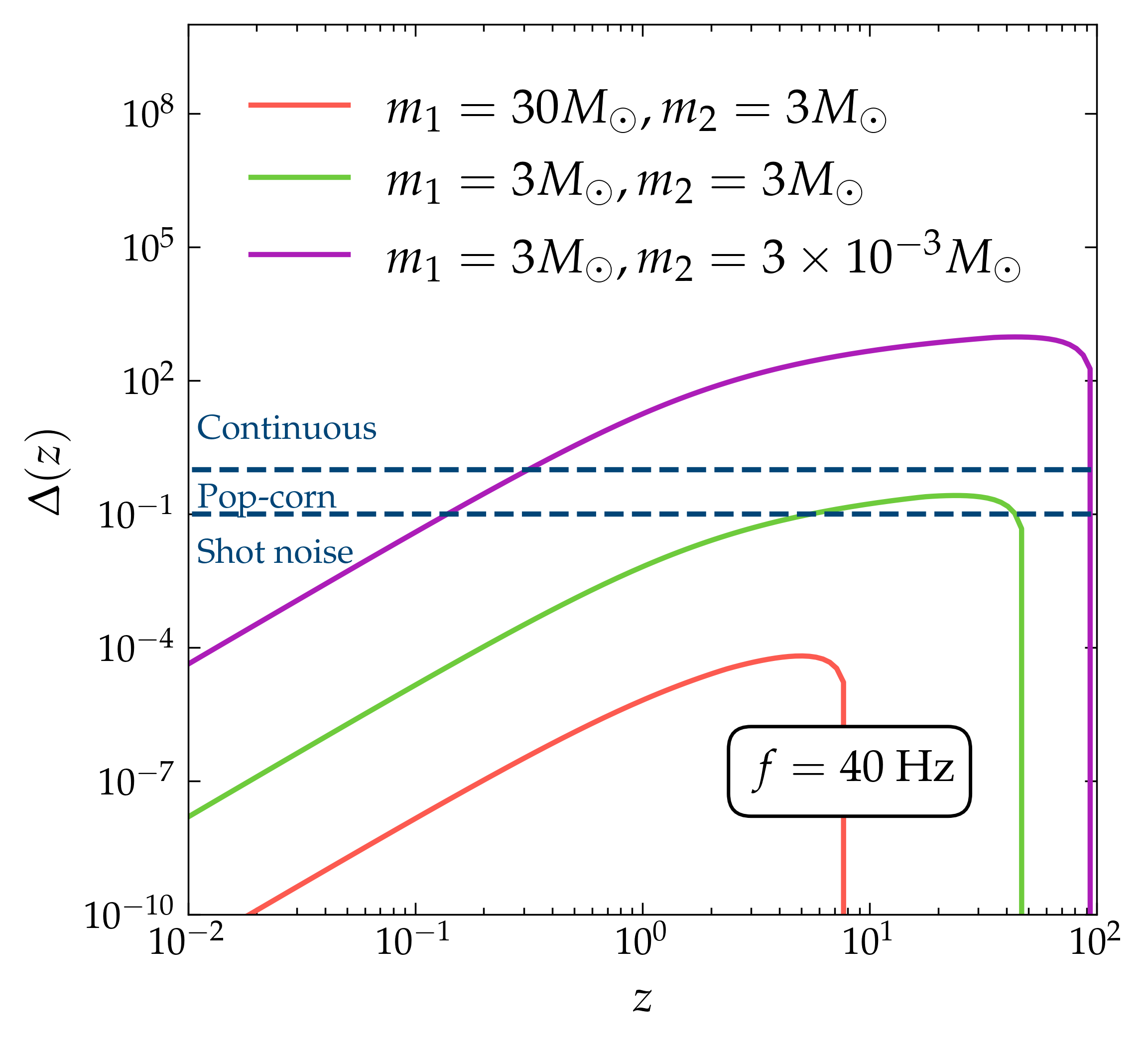}
  \caption{The duty cycle as a function of the maximum redshift for three pairs of component masses $m_1$ and $m_2$ of early PBH binaries, in one e-fold of frequency around 40~Hz. The popcorn region is enclosed by the two dashed lines indicating $\Delta(z) = 0.1$ and $1$. }
  \label{fig:Delta_of_z}
\end{minipage}
\begin{minipage}[t]{0.99\columnwidth}
  \includegraphics[width=1.\linewidth]{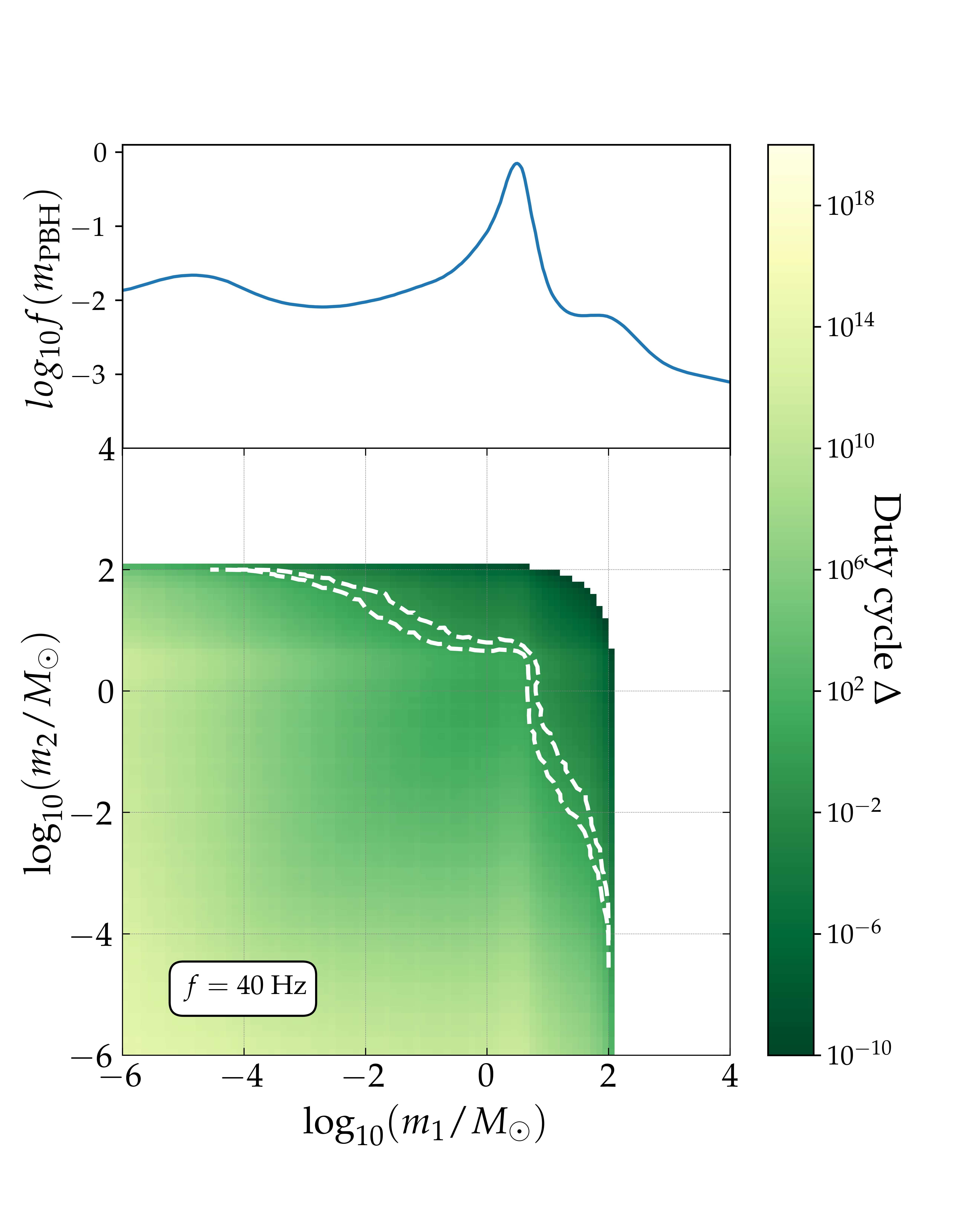}
 \caption{Top panel: Assumed PBH mass distribution, from~\cite{Bagui:2021dqi}, showing features from the sound speed reduction at the electroweak and QCD epochs.  Bottom panel: The duty cycle distribution as a function of the masses $m_1$ and $m_2$ for early PBH binaries with the considered mass distribution,  a maximum redshift $z_{\rm max} = 100$, at a frequency of 40~Hz. The white dashed lines enclose the popcorn region between $\Delta = 0.1$ and $1$. }
  \label{fig:Delta_of_m}
\end{minipage}
\end{figure}

Based on the magnitude of the duty cycle, we define three different regimes: 
\begin{itemize}
    \item $\Delta > 1$, where the overlap of the signals generates a \textit{continuous} Gaussian background; 
    \item $\Delta < 0.1 $, where long periods of silence separate two signals, resulting in a \textit{shot-noise} background; 
    \item $0.1 < \Delta < 1$, where the signals may overlap, but their statistics are no longer Gaussian, giving rise to a \textit{popcorn} background.
\end{itemize}
Fig.~\ref{fig:Popcorn_plot} shows the typical time-domain behaviour of the strain signal in each of these three regimes using a simple toy model for the population.

By inserting Eqs.~\eqref{eq:cosmomerg} and~\eqref{eq:duration} into Eq.~\eqref{eq:duty_cycle_definition}, we calculate the duty cycle of early PBH binaries as a function of the component masses $m_1$, $m_2$ and of the maximum redshift $z$ for the considered population of early PBH binaries.  Some representative results are shown in Fig.~\ref{fig:Delta_of_z} for a frequency bin of one $e$-fold at 40~Hz (corresponding to the maximum LVK GWB sensitivity).
We show three pairs of masses that correspond to binaries combining a $3 M_\odot$ PBH from the QCD peak in the mass function (see Fig.~\ref{fig:Delta_of_m}) with one ten times heavier, of equal mass, or 1000 times lighter.  In the case of equal-mass binaries from the peak, the popcorn regime is reached for sources at high redshift, typically at $z\gtrsim 4$, that are too faint to be individually resolved. Binaries with a heavier PBH only produce signals in the shot noise regime, while those with a lighter PBH produce a shot-noise signal only at low redshift.  Above some redshift value that depends on the masses, the duty cycle is strongly damped because the redshifted ISCO frequency goes out of the considered band.
For asymmetric binaries with $m_2 \lesssim M_\odot$, the background is continuous at high redshift due to the much longer signal duration. 
In our analysis below, we  use a redshift cutoff $z_\mathrm{max}=100$.

The duty cycle distribution in the $(m_1,m_2)$ plane is shown in Fig.~\ref{fig:Delta_of_m} for a frequency at 40~Hz.  As expected, there is no signal when $m_1+m_2$ leads to $f_{\rm ISCO} /(1+z_{\rm max})> 40$~Hz.  Depending on their mass, PBHs can lead to either a shot-noise, popcorn or continuous GWB.  Binaries involving at least one strongly subsolar-mass PBH typically produce a continuous signal, due to their enhanced merger rates and signal duration.  When the two masses are above a few solar masses, one gets a  shot noise signal.  The popcorn region takes the form of a band, including binaries with two black holes from the peak and extended to asymmetric binaries combining a subsolar-mass black hole with one larger than a few solar masses.

\textbf{Astrophysical sources---}
In order to compute the GWB from black holes of stellar origin, we use the fiducial \textsc{power law + peak} (PP) model for the  differential merger rate as a function of the primary mass $m_1$, ${\rm d}\Bar{R}/{\rm d}m_1$, and the mass ratio $q = m_2/m_1$, ${\rm d} \Bar{R}/ {\rm d} q$, with best-fit parameter values coming from the population analysis of the GWTC-3 catalog~\cite{GWTC-3-pop-LIGOScientific:2021psn}. The PP model is a combination of a truncated power law with sharp cutoffs at 5 and 80~$M_\odot$ and a Gaussian peak centred at $33 M_\odot$~\cite{GWTC-3-pop-LIGOScientific:2021psn}. This particular model is chosen because it provides the best Bayes factor after GWTC-2~\cite{GWTC-2-pop-LIGOScientific:2020kqk}. The redshift dependence of the rate follows the star formation rate (SFR) density $\psi(z)$ from \cite{CBC_merger_rates_Santoliquido:2020axb, CBC_merger_rates_Madau:2016jbv},
\begin{equation}
\psi(z) \propto \frac{(1+z)^{2.6}}{1 + \left(\frac{1+z}{3.2}\right)^{6.2}},
  \label{eq:psi}
\end{equation}
which is normalised such that $\psi(0)=1$.
The differential merger rate per logarithmic mass interval is thus given by $
{\mathcal R} (m_1, m_2, z) = (1/2) \left[{\rm d}\Bar{R} / {\rm d} m_1 \right] \thinspace \left[{\rm d}\Bar{R} /{\rm d} q \right] \thinspace m_2 \thinspace \psi(z)$, where the factor $1/2$ enforces the mass ordering $m_1 > m_2$ used by the LVK collaboration.

For binary neutron stars (BNS), we model the merger rates with a flat mass distribution between $1.2 $ and $2.0 M_\odot$, as suggested by the GWTC-3 population analysis, with a total BNS rate in this range of $98 \,{\rm yr}^{-1} \, {\rm Gpc}^{-3}$, i.e. the mean value obtained 
for the binned Gaussian process model 
in the NS mass range~\cite{GWTC-3-pop-LIGOScientific:2021psn}. The redshift dependence is assumed to be the same as that for astrophysical BHs. 

\begin{figure*}
    \centering
\includegraphics[width=0.45\linewidth]{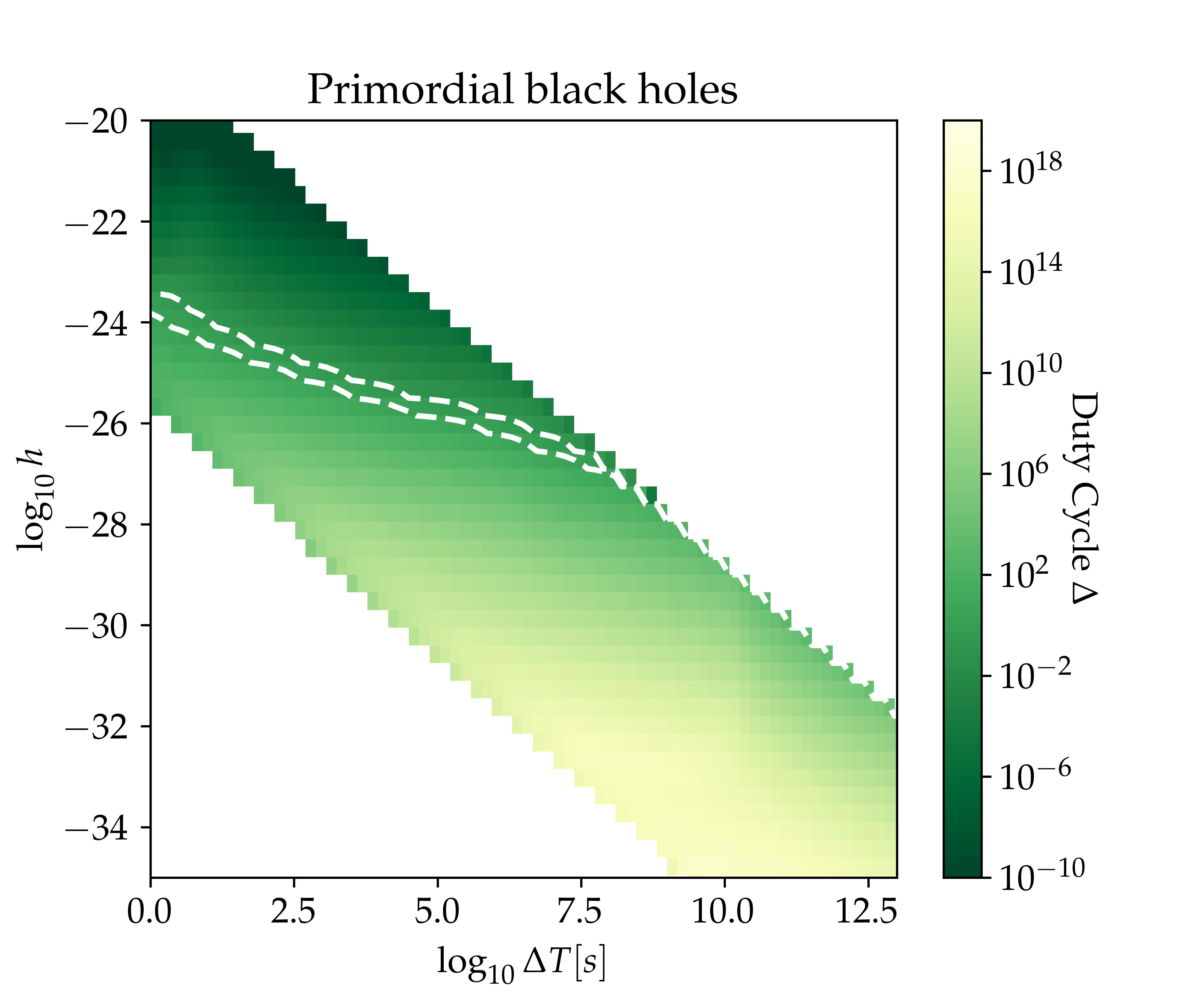}
\includegraphics[width=0.45\linewidth]{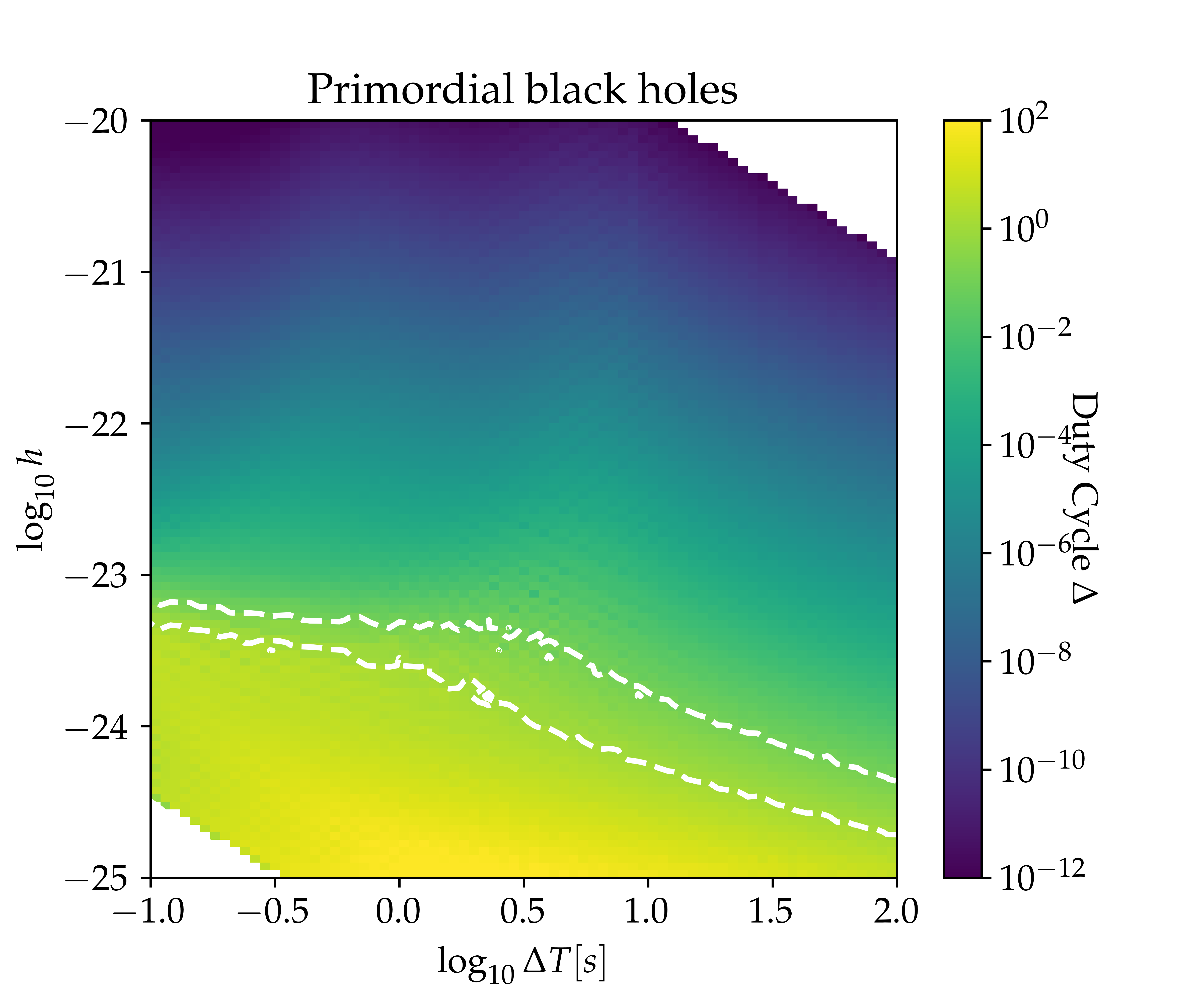}  \\
\includegraphics[width=0.45\linewidth]{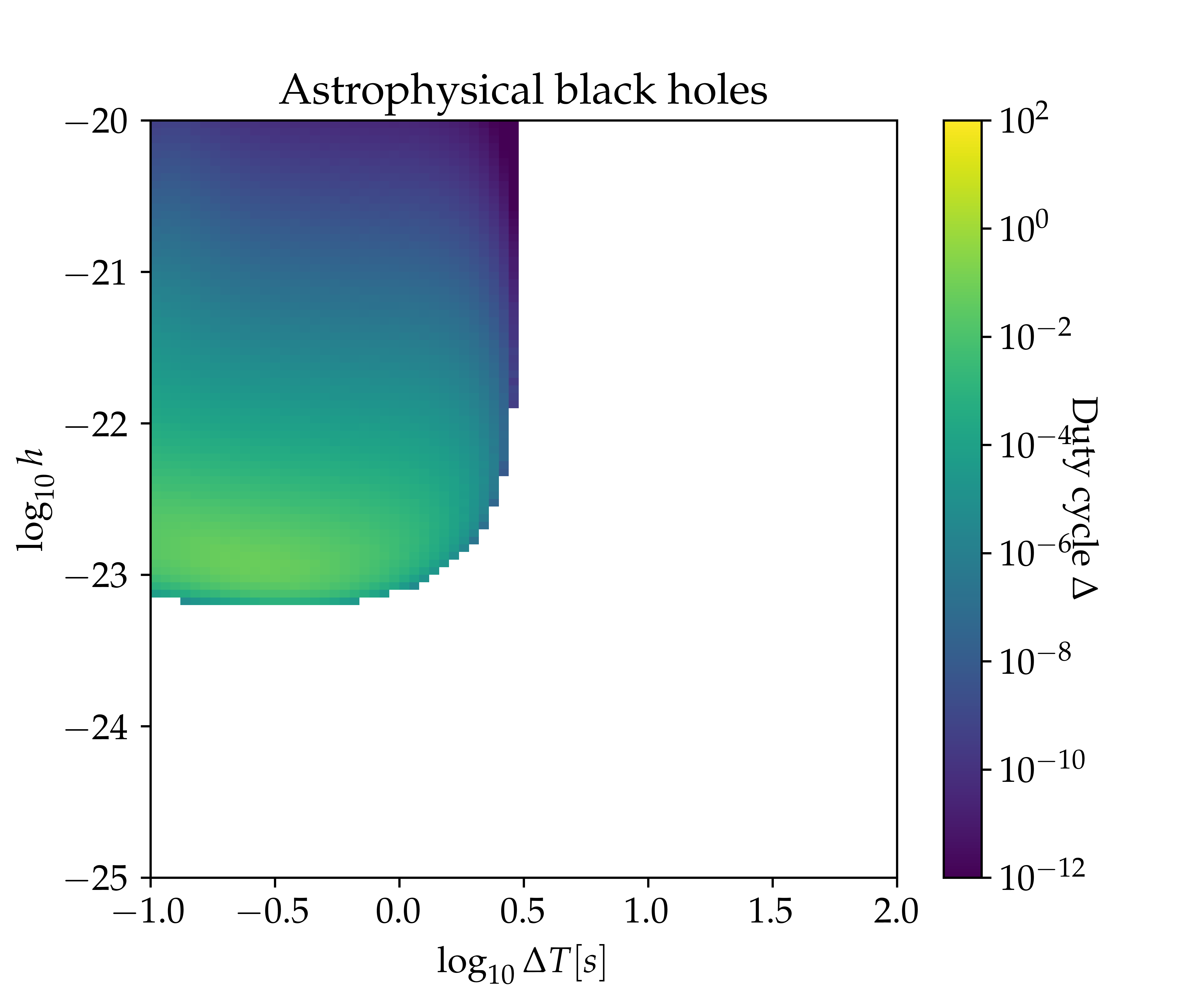}
\includegraphics[width=0.45\linewidth]{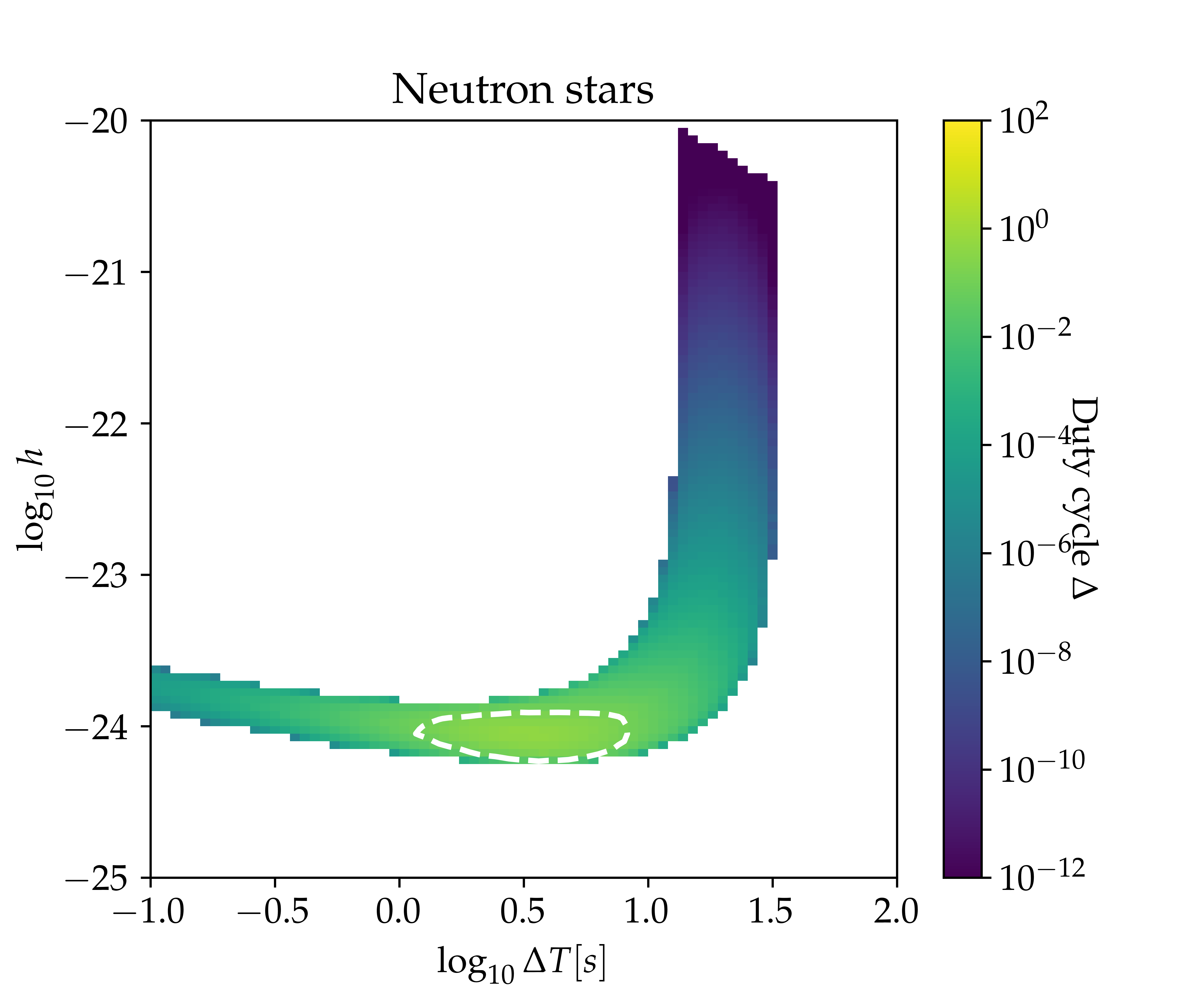}
    \caption{%
    Expected distributions of the duty cycle with GW signal strain $h$ and duration $\Delta T$ at $f=40$~Hz, for our benchmark PBH model (two top and  panels) and for astrophysical black holes (third panel) and neutron stars (bottom panel). The white contours indicate the popcorn regions, defined as $0.1<\Delta < 1$.}
    \label{fig:duty_cycle_dists}
\end{figure*}

\textbf{Duty cycle distributions---}
As one can see from Eq.~\ref{eq:duty_cycle_definition}, the measurement of the \textit{total} duty cycle by an \textit{ideal} detector only gives access to the integrated merger rates over component masses and redshifts, which does not allow one to distinguish two merger rate distributions producing the same total rate. In practice, the situation is further complicated by the presence of detector noise, which prevents one from resolving many of the sources that contribute to the total duty cycle.
Instead, one can only measure the duty cycle of sources with GW strains that exceed some detection threshold, which itself will depend on the signal duration in the considered frequency band. The total measured duty cycle is therefore a bad discriminator between models in practice.

In the case of PBHs with broad mass distributions, the total duty cycle is always much larger than that associated with resolvable sources in the LVK frequency range, due to a very large number of light binaries that emit GWs far below the detection threshold. It is therefore crucial to go beyond the simple metrics described above, and to design search pipelines that can reconstruct or constrain the merger rate distribution in order to distinguish between the possible origins of the compact objects.

Instead of measuring the total duty cycle as a function of the frequency, we stress the need to design methods to reconstruct and constrain the duty cycle \textit{distribution} associated with different signal durations, GW strain amplitudes and frequencies, i.e. $\dd{\Delta} / (\dd{\ln T} \dd {\ln h} \dd{\ln f})$.  We point out that this distribution of the duty cycle is uniquely defined for a merger rate distribution as a function of the chirp masses, mass ratios (or equivalently of $m_1,\, m_2$) and redshifts.  Indeed, by examining the Eqs.~\ref{eq:duty_cycle_definition} and~\ref{eq:duration} , one sees that by probing the four independent quantities $\Delta$, $T$, $f$ and the single-source strain $h$ given by
\begin{equation}
h=\frac{4}{D_{\rm L}} \left(\frac{G \mathcal{M}_c}{c^2} \right)^{5/3} \left( \frac{\pi f^{\rm obs}}{c} \right)^{2/3}, 
\end{equation}
where $\mathcal M_c$ and $f^{\rm obs}$ are the redshifted mass and frequency in the observer's frame, one can access the four other independent quantities: the rates $R^0$, the masses $m_1$ and $m_2$ and the redshifts $z$.
If a model for the merger rate is specified (e.g. the model for early PBH binaries in Eq.~\ref{eq:cosmomerg}), one can then reconstruct or constrain the compact binary mass distribution, which would be highly valuable in the study of PBH scenarios.  

We therefore propose the development of a new analysis pipeline to extract these quantities from time-frequency data.  The Generalized Frequency-Hough transform could be appropriate to identify coincident transient continuous-wave sources in detectors and to probe their duty cycle.  This method has already been used on LVK data to search for light subsolar-mass PBHs~\cite{Miller:2024fpo,Miller:2024jpo} with $\Delta T > 1$ hour, with a strain sensitivity down to $10^{-25}$ for persistent sources, but the search of shorter signals is computationally challenging.  It could be combined with discrete wavelet transform algorithms to cross-correlate cleaned time-frequency data between different detectors and extract or constrain signals with durations $\Delta T$ going down to $0.1$ s, exploiting the downsampling properties of wavelet algorithms~\cite{Klimenko:2004qh} (see also ~\cite{Chassande-Mottin:2017qim,Mathur:2025ogv,Taylor:2024cgu,Roy:2022teu}).

In order to demonstrate the differences between our PBH model and the astrophysical channel in terms of this proposed observable, we transform our distributions of the duty cycle as a function of $m_1,\, m_2$ into duty cycle distributions as functions of $\Delta T$ and $h$.  The obtained distributions for PBHs, ABHs and NS are shown in Fig.~\ref{fig:duty_cycle_dists}, where the popcorn and continuous regions are contoured, for a frequency $f=40$~Hz.  For PBHs, the continuous regime extends up to $\Delta T$ values larger than the typical duration of an observing run, but the strain amplitudes associated with the corresponding individual sources in this regime are far below the detection threshold. This is due to the presence of numerous PBH binaries much lighter than the solar mass.
The strain associated with the popcorn region peaks around $h\sim 10^{-23}$ for $\Delta T$ between 0.1 and 100~s, with PBHs, astrophysical BHs and neutron stars all having distinct features in this regime.  For PBHs, one obtains a thin band of roughly constant strain amplitude across a broad range of signal durations in which the signal is popcorn-like.  For astrophysical BHs, the popcorn region is absent and the GWB is of shot-noise type and restricted to $\Delta T \lesssim 3$ s, due to the $5 M_\odot$ mass cutoff in this population. For neutron stars the population extends to lower masses, allowing $\Delta T$ to reach about 30 seconds, with a popcorn region that delimits a peak in the duty cycle distribution and the continuous regime being  localized in a relatively thin band, contrary to the PBH case.  The shape of the popcorn, shot-noise and continuous regions is therefore a discriminator between the three channels, even if one does not confidently detect individual subsolar-mass black hole mergers with standard matched-filtering techniques.

\textbf{Conclusion and discussion---}  Stellar-mass PBH mergers are challenging to distinguish from astrophysical black-hole or neutron-star mergers in LVK observations with standard matched-filtering or GWB searches.  In this \emph{Letter}, we have argued that the time-domain properties of the GWB such as the duty cycle can in principle be used as a discriminator between these populations, and have presented the first calculation of the expected duty cycle for early PBH binaries with a realistic extended mass distribution.
Our findings motivate the development of new search pipelines to constrain or reconstruct the duty cycle distribution as a function of the individual signal duration, frequency, and amplitude, which could be then translated into constraints on the merger rate distribution as a function of the component masses and redshift.  

We have calculated this duty cycle distribution for a representative PBH scenario, and have found this to be a robust discriminator between PBHs, astrophysical black holes and neutron stars.  The shape of the ``popcorn'' regime, in which the duty cycle is slightly below unity, is of particular interest.  For PBHs, this popcorn regime at a frequency of $40$ Hz   corresponds to a particular range of strain amplitudes and 
is almost independent of the signal duration. In contrast, for astrophysical black hole and neutron star mergers this regime either does not exist or corresponds to an isolated peak in the duty cycle distribution. 
If a GWB is observed by upcoming LVK observing runs or by next-generation ground-based observatories such as Einstein Telescope, probing the duty cycle distribution and identifying the properties of the popcorn region will allow one to distinguish the origin of this GWB. For PBH binaries, this observable could even be used to constrain the PBH merger rate and the mass distribution down to the planetary-mass scale, shedding light on possible PBH formation scenarios. 

Our work strongly motivates the development of new analysis pipelines to constrain this duty cycle distribution. Such a pipeline could for instance rely on existing generalized Frequency-Hough analyses \cite{Miller:2020kmv, Miller:2024jpo, Miller:2024fpo, LIGOScientific:2025vwc}, which have so far only been used to search for at least hour-long individual signals from subsolar-mass PBHs, but not for their popcorn GWB. There is also scope for improvements to the theoretical modelling presented here, for example by developing more sophisticated population and merger-rate models (for instance of the suppression factor $f_\mathrm{sup}$ that plays a critical role in determining GWBs), or by including the merger and ringdown contributions to the binary GW emission (although these improvements are unlikely to drastically modify the results presented here).
Finally, the duty cycle distribution could also be calculated at nanohertz frequencies for PBH binaries formed in clusters that could be relevant for pulsar timing arrays, in order to identify the type of GWB one would expect in that band and discriminate between possible origins.

\begin{acknowledgments}
    \textbf{Acknowledgments---}
    This material is based upon work supported by NSF’s LIGO Laboratory which is a major facility fully funded by the National Science Foundation. The LIGO document number of this article is LIGO-P2500404.
    The authors are grateful to Joseph D. Romano for carefully reading the manuscript and providing useful feedback.
    E.B. and S.C. are suppported by an IISN Grant of the F.R.S.-FNRS.
    F.D.L. was supported by a FRIA (Fonds pour la formation à la Recherche dans l’Industrie et dans l’Agriculture)
    Grant of the Belgian Fund for Research, F.R.S.-FNRS
    (Fonds de la Recherche Scientifique-FNRS) in the early phases of this work. 
    A.C.J was supported by the UK Engineering and Physical Sciences Research Council (EPSRC) through a Stephen Hawking Fellowship (grant number EP/U536684/1), and by a Gavin Boyle Fellowship at the Kavli Institute for Cosmology, Cambridge. 
\end{acknowledgments}

\bibliography{PBH_biblio.bib}

\clearpage
\pagebreak
\appendix
\section*{END MATTER}

\textbf{The popcorn GW background---}
When considering a population of compact binary coalescences (CBCs), the superposition of GWs from individual binaries generates a GWB characterized by $\Omega_{{\rm{GW}}}$, the relative GW energy density $\rho_{\rm GW}$ to the critical density $\rho_{\rm c}$ today, per unit of observed logarithmic frequency.  For a single binary, the radiated energy is related to the strain amplitude as $\rho_{\rm GW} =  \pi h^2 f^2/ (4 G)$.  The GW energy density spectrum sourced by CBCs can therefore be expressed as
\begin{align} \label{eq:background}
\Omega_\mathrm{GW}(f) = \frac{\pi f^2}{4 \pi \rho_{\rm c}} \int {\rm d } z \, \Delta(z,f) h(f)^2~.
\end{align}
In order to partition this spectrum into shot-noise, popcorn and continuous contributions, we sort $h(f)$ in each mass and redshift bin, from the highest to lowest value. The duty cycle in each bin is then summed until the total duty cycle gives $\Delta = 0.1$ and the corresponding shot-noise GWB $\Omega_{\rm GW}^{\rm shot-noise}$ is obtained with Eq.~\ref{eq:background}, where integrals becomes a sum over those bins.   The popcorn GWB $\Omega_{\rm GW}^{\rm popcorn}$ is obtained by summing subsequent bins, until a total duty cycle $\Delta = 1$ is reached, while the continuous GWB $\Omega_{\rm GW}^{\rm continuous}$ is obtained by summing over the remaining bins.  
Other prescriptions are possible, e.g. by sorting bins in terms of signal-to-noise ratio in the detector, giving explicitly detector-dependent results.

\begin{figure}[b]
 \centering
  \includegraphics[scale=0.73]{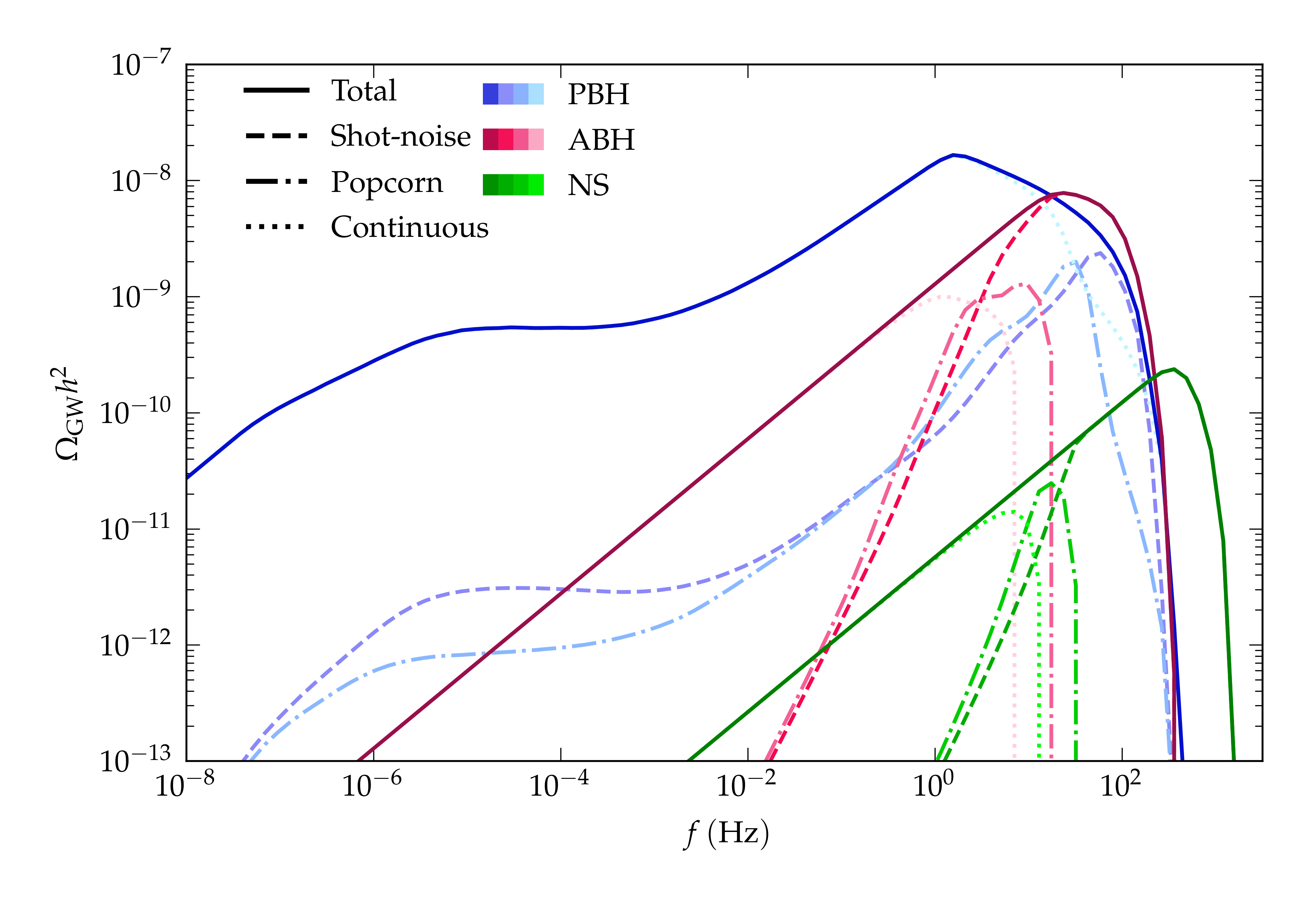}
  \caption{%
  Comparison between the GWB from early PBH binaries with the mass distribution shown in Fig.~\ref{fig:Delta_of_m} (blue lines) and the one from ABHs (red and pink lines) and NS (green lines).  The solid, dashed, dot-dashed and dotted lines show respectively the total background and the shot-noise, popcorn and continuous distributions, obtained when sources are ranked with respected to the strain $h(f)$.  For PBHs, the popcorn background is dominant between 10 and 100~Hz in the sensitivity band of LVK, whereas for ABH and NS, the shot-noise GWB dominates.}
  \label{fig:Astro}
\end{figure}

These GWBs for PBHs, ABHs and NS are shown in Fig.~\ref{fig:Astro}.  For PBHs, at frequencies below 10~Hz, the GWB is mostly continuous. Between 10~Hz and 100~Hz, where LVK is the most sensitive to the GWB, the three regimes contribute at a similar level, with a peak for the popcorn GWB around 40 Hz, whereas for ABHs and NS the main contribution comes from the shot noise regime.  At higher frequencies, the shot-noise regime is dominant in all cases.  This particularity for PBHs comes from the existence of binaries with at least one subsolar-mass component, and from higher redshifts, which increases the duty cycle compared to ABHs and NS.

These findings provide a clear distinction between the GWBs, showing that the duty cycle can in principle be used to distinguish astrophysical binaries from PBH binaries as GWB sources in the LVK band.  The GWB from PBHs with an extended mass function could be near the current limits, therefore for upcoming observing runs, this demonstrates the interest of developing new and specific search methods based on the duty cycle~\cite{Drasco:2002yd, Smith:2017vfk, Lawrence:2023buo, Kou:2025bhk} to improve the sensitivity to the popcorn GWB, given that the standard cross-correlation search performed in LVK is based on a continuous-Gaussian signal model that does not take into account the intermittent nature of the signal. This approach would also be able to distinguish other cosmological GWB signals that have been proposed in the literature, as these would be purely continuous, with no popcorn or shot-noise component whatsoever.

One can compare our results to those of~\cite{Braglia:2022icu} obtained for late PBH binaries in clusters, which typically lead to a subdominant GWB compared to early binaries for broad mass distributions and to lower duty cycles for binaries with subsolar-mass PBHs.

\end{document}